\documentstyle[12pt,epsf]{article}

\newcommand{\be}{\begin{equation}}
\newcommand{\ee}{\end{equation}}
\newcommand{\bea}{\begin{eqnarray}}
\newcommand{\eea}{\end{eqnarray}}
\newcommand{\half}{{1 \over 2}}
\newcommand{\pq}{$(p,q)$ }

\def\href#1#2{#2}

\def\IZ{\relax\ifmmode\mathchoice
{\hbox{\cmss Z\kern-.4em Z}}{\hbox{\cmss Z\kern-.4em Z}}
{\lower.9pt\hbox{\cmsss Z\kern-.4em Z}}
{\lower1.2pt\hbox{\cmsss Z\kern-.4em Z}}\else{\cmss Z\kern-.4em
Z}\fi}
\def\IR{\relax{\rm I\kern-.18em R}}
\def\IC{\bf C}
\font\cmss=cmss10 \font\cmsss=cmss10 at 7pt

\begin{document}
\begin{titlepage}

\rightline{SU-ITP-97-53}
\rightline{hep-th/9801067}
\rightline{January 12,1998}
\rightline{v2: \today}
\begin{center}

{\Large\bf BPS Spectrum of 5 Dimensional Field Theories, (p,q) Webs \\ and Curve
Counting}

\vskip 0.5cm

{\bf Barak Kol} \\
\tt{barak@leland.stanford.edu} \\
\vskip 0.3cm
{\bf J. Rahmfeld} \\
\tt{rahmfeld@leland.stanford.edu} \\
\vskip 0.5cm
Department of Physics\\Stanford University\\Stanford, CA 94305, USA

\end{center}

\begin{abstract}
We study the BPS spectrum of supersymmetric 5 dimensional field theories and
their representations as string webs. It is found that a state of given charges
exists when it has a representation as an irreducible string web. Its spin is
determined by the string web. The number of fermionic zero modes is $8g+4b$,
where $g$ is the number of internal faces and $b$ is the number of boundaries.
In the lift to M theory of 4d field theories such states are described by
membranes ending on the 5-brane, breaking SUSY from 8 to 4, and $g$ becomes the
genus of the membrane. Mathematically, we obtain a diagrammatic method to find
the spectrum of curves on a toric complex surface, and the number of their
moduli.
\end{abstract}

\end{titlepage}
\newpage


\section{Introduction}
In \cite{AHK} $(p,q)$ webs of 5-branes were shown to serve as a
model for (some) 5d supersymmetric field theories. The definitions and relations
are reviewed in section \ref{WebRev}. In that paper some open questions were
mentioned, and the first two were:
\begin{enumerate}
\item 5d field theories can be constructed both from brane configurations (webs)
 \cite{HW,AH} and by compactifying M theory on a Calabi-Yau (CY)
  threefold while decoupling gravity by shrinking a 4 cycle. It was
conjectured that there is a mapping between webs and Calabi-Yau manifolds.
\item The BPS spectrum was shown to be encoded as \pq string webs inside the 5-brane web.
It was not known how to determine the multiplet type (hyper,vector..) of a
state, and how to determine whether a marginally bound state exists.
\end{enumerate}

The first question was answered by Leung and Vafa \cite{LV} showing how the
$(p,q)$ web is related to toric manifolds. We review that connection in section
\ref{toric}. In this paper we address the second question and study the BPS
spectrum. Consider a string web that has boundaries on a 5-brane web. One has to
find the zero modes, and look for the ground states of the quantum mechanics
which they define. The transformation properties of these ground states under
the super-Poincare group give the multiplet type of the particle.

We find that
\begin{enumerate}
\item  The ground state of the string web is indeed supersymmetric.
\item A marginally bound state exists whenever the string web is irreducible, as anticipated in
\cite{AHK}, namely when a generic web cannot be decomposed into sub-webs. Thus
we find sites on the charge lattice which are occupied by marginally bound
states, but it is not clear whether, in general, other sites are occupied.
\item  The multiplet type of a string web with no internal faces (corresponding to a zero genus curve)
is determined. Recall that a massive BPS state in 5d transforms under
$SO(4)=SU(2)_p \times SU(2)_b$, where the subscripts $p,b$ refer to preserved
and broken supersymmetries. The particle transforms as $ (j)_p \otimes H_0$,
where $H_0=2(0)_b \oplus (1/2)_b$ is a half hypermultiplet, generated by the
broken supersymmetries. The quantization of a web might result in a number of
such representations. The highest spin representation, $j$, is
\be
2j = n_X-1
\ee
where $n_X$ is the number of external legs. $j=0$ is a hyper and $j=1/2$ is a
vector.
\item For a general web we find the zero modes, though presently we do not solve the
quantum mechanics. The number of bosonic zero modes (ignoring the 4 translation
modes) is $n_{BZM}=n_X + F_{int}-1=F-1$, where $F_{int}$ in the number of
internal faces, and $F$ is the total number of faces. In case there are "hidden"
faces, $F$ should be replaced by the total number of points in the respective
grid diagram \cite{AHK}. Comparison with the geometrical analysis shows that
these moduli are complexfied. The bosonic zero modes define the moduli space for
the web, which is the target for the quantum mechanics. The number of fermionic
zero modes \cite{BergmanKol}, and hence the maximum spin in a reduction to 4d,
$j_4$, are
\bea
n_{FZM} = 8F_{int}+4n_X \nonumber \\
 \left| j_4 \right| \leq n_X/2+F_{int}
\eea
After lifting to M theory we find the number of fermionic zero modes for a
massive membrane ending on a 5-brane and breaking 8 supercharges to 4
\be
n_{FZM}=8g+4b
\ee
generalizing the results of \cite{HennYi,Mikhailov,HennRev}, where $b=n_X$ is
the number of boundaries and $g=F_{int}$ is the genus.
\end{enumerate}

The method is to use \cite{LV} to translate webs into geometry, and then borrow
the results of Witten \cite{Witten_phase}. The number of fermionic zero modes is
found by solving the zero modes equation, or by operating with locally preserved
supersymmetries  on the bosonic zero modes \cite{BergmanKol}. The argument is
presented in section \ref{BPSspec}. Three examples are worked out in section
\ref{examples}.

Mathematically, we count curves \cite{KMV,MNW9705} on a toric complex surface (4
real dimensions), or a few intersecting toric surfaces. We find a diagrammatic
method to determine which homology classes are occupied by curves, and the
number of their moduli.

We would like to point out directions for further research:
\begin{itemize}
\item We determined so far the zero modes but did not solve for the ground states of the quantum mechanics.
The moduli space contains singularities, in which webs or curves are reducible,
which may affect the quantum mechanics. In particular, for the zero genus case
we know the highest spin multiplet, but others might arise at the singularities.
\item It is known that irreducibility is a sufficient condition for a bound
state. What is the necessary condition?
\item It was shown that webs can describe toric geometries. It is still unknown whether there are
 brane configurations for other geometries, such as ${\bf B}_k$ for $3<k \le 8$.
\end{itemize}

Version 2: The counting of fermionic zero modes is updated due to
\cite{BergmanKol}, which found some zero modes that were ignored in the previous
literature (see introduction). Consequently, we do not know the general solution
of the quantum mechanics on the moduli of the web (except for zero genus). The
list of directions for further research is shortened since some points were
clarified in the meantime: The (real) moduli of the web indeed map to moduli of
the lifted membrane. The periodic string webs which were described by Sen
\cite{SenNet} (see also
\cite{MatsuoOk,ReyJunc,KroghLee,MarkSmolin,KishSasa,BergWeb16,GZ}), are studied
in \cite{BergmanKol} and their fermionic zero modes are determined. The zero
modes counting seems to be consistent with the index theorem of the 2d Dirac
operator.

\section{Review of \pq webs}
\label{WebRev}

Five dimensional field theories were thought to be non-existent since they are
not renormalizable. However, 5d conformal theories enable the definition of UV
fixed points, which can be perturbed in the IR, for example by Yang-Mills terms.
The study of 5d, supersymmetric $N=1$ (8 supercharges) field theories was
initiated by Seiberg \cite{Seiberg5d} and developed in
\cite{MS,DKV,GMS,IMS,DFKS}. These papers took both a field theoretic approach
and a geometric approach considering M theory on a CY. It was found that the
theory is characterized by a prepotential which is cubic in the moduli. The
first derivatives give the tension of BPS monopolic strings and the second
derivatives give the running coupling constant (the metric on moduli space).
There exists a quantized parameter with a finite number of allowed values,
called $c$, which is related to the coefficient of the Chern-Simons term.

\begin{figure}
\centerline{\epsfxsize=100mm\epsfbox{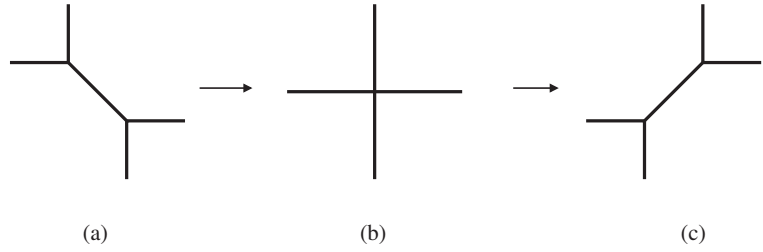}}
\medskip
\caption{A flop transition.}
\label{flop}
\end{figure}

\begin{figure}
\centerline{\epsfxsize=100mm\epsfbox{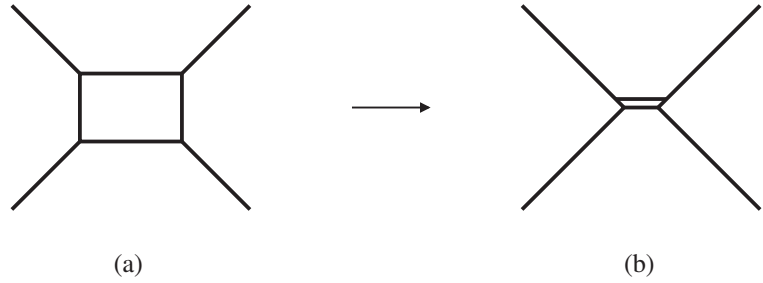}}
\medskip
\caption{An enhanced gauge symmetry singularity.}
\label{Wsing}
\end{figure}

In \cite{AHK} $(p,q)$ webs of 5-branes were shown to serve as a model for (some)
5d supersymmetric field theories. A \underline{$(p,q)$ web} is defined by a
collection of edges (finite or semi-infinite) in the $(x,y)$ plane. Each edge is
carrying a \pq label, where \pq are relatively prime integers. The slopes are
constrained according to
\be
\Delta x : \Delta y = p : q.
\label{slope}
\ee
The edges are allowed to meet at vertices provided that the \pq charge is
conserved
\be
{\sum_{i}^{}{\left[\matrix{ p \cr q \cr }\right]_i}} = 0.
\label{vertex}
\ee
The slope condition (\ref{slope}) allows a sign ambiguity for \pq, which is
fixed by a choice of orientation. To check the vertex condition (\ref{vertex})
the edges should be oriented to be all incoming, or all outgoing.
 Assigning to a \pq  edge a tension
\be
T_{p,q}=\sqrt{p^2+q^2}
\label{tension}
\ee
these conditions ensure the equilibrium of forces at a vertex. See figures
(\ref{flop},\ref{Wsing}) and others for examples. The definition of the web
\cite{AH,AHK} developed gradually. It originated with the work of Schwarz
\cite{SchwarzSL2z} on
\pq strings. The vertex condition was found by Aharony, Sonnenschein and Yankielowicz
\cite{ASY}. The BPS nature of a vertex at mechanical equilibrium was discussed
by Schwarz \cite{SchwarzTASI}, and finally the full definition including the
slope condition was given by Aharony and Hanany \cite{AH}. A proof that a string
web preserves $1/4$ of the supercharges was given in
\cite{DasguptaMukhi,MatsuoOk}.

String theory has two objects that can realize a \pq web: the \pq strings and
the \pq 5-branes of type IIB. Employing the \pq 5-branes we can take the other
$4+1$ dimensions of the 5-brane to be common, and get as the low energy theory
of the web a 5d $N=1$ field theory \cite{AH}. In accordance with the general
construction of brane configurations \cite{HW}, this requires the energies on
the brane to be much smaller than the Planck mass so that gravity decouples. For
a general complex scalar of IIB, $\tau=\chi/2\pi+i/\lambda$, where $\chi$ is the
RR scalar and $\lambda$ is the IIB string coupling, the plane of the web would
be related to the physical plane of the 5-brane through a linear transformation.
In \cite{AHK} it was conjectured that $\tau$ is a redundant parameter for the
low energy field theory and can be set to the S self-dual point $\tau=i$, so we
can identify the two planes.
 It was found that for many field theories (at least for those that
are related to toric spaces) the vacuum structure could be read from the web
\cite{AHK}:

\begin{tabular}{l l}
\underline{Field theory} & \underline{Web} \\
$\bullet$ Local $U(1)$ gauge symmetries & Local deformations of the web \\
 $\bullet$ Global U(1) symmetries          & Global deformations \\
  $\bullet$ Tension of monopolic string & Area of face \\
 $\bullet$ Running coupling constant      & Mass of a local deformation mode \\
  &(roughly the perimeter of the face) \\
\end{tabular}

These quantities allow to determine the prepotential of the theory.

The BPS spectrum was found to be described in terms of strings ending on the
5-brane \cite{AHK}. In addition to the strings, \underline{''strips"} or strings
inside the 5-brane are a second building block. Strips differ from strings by
their tension, which is (for $\tau=i$)
\bea
T_{p,q}^{strip}=1/\sqrt(p^2+q^2).
\eea
\underline{Webs of strings and strips} can be built according to the following rules:
\begin{itemize}
\item String boundary: a \pq string can only end on a \pq 5-brane.
\item (String) slope: $\Delta x : \Delta y = -q : p$. This assures there is \underline{no parallel
force} to the 5-brane.
\item String vertex. The \pq charge is conserved at string vertices.
\item Strip type. A \pq strip lies inside a \pq 5-brane.
\item String-strip vertex - \underline{''bend"}. Strips are allowed to bend out of the
5-brane to become strings. Given an integer vector \pq, there is an integer
vector $(e,f)$ that completes it to a basis of $\IZ ^2$. A collection of $a$
\pq strips are allowed to bend into a $a(e,f)+b(p,q)$ string for arbitrary $b$. Note
that the no-parallel-force rule is satisfied. If $a,b$ are not relatively prime
this configuration is reducible to a multiple of bend configurations.
\item Strip boundary. Strips end at vertices of the 5-brane web, but they
cannot end on any vertex. A strip should be thought as ending on one of the
other 5-brane edges coming out of a vertex. The sum of all
\pq charges of boundaries is required to vanish. For example, a strip
is allowed along a 5-brane edge iff there are parallel and opposite 5-branes
coming out of the two vertices. This would be a hyper.
\end{itemize}
Examples of string webs are shown in figures
(\ref{E0},\ref{E0-2},\ref{E1},\ref{E1tilde}).

In M theory these configurations are known to be lifted to membranes ending on
the M5-brane. The membranes are holomorphic in a complex structure that is
orthogonal to the 5-brane complex structure.

At this point we can give a specific example for the question that is discussed
in this paper. Consider the pure $SU(2)$ theory (see figure \ref{E1}a). It has
two simple BPS states. The W, represented by the vertical line, has charge
$(n_e,I)=(1,0)$, where $n_e$ is the electric charge, and $I$ is the instanton
charge. The instanton, represented by the horizontal line has charge $(1,1)$.
The type of question we would like to answer in this paper is whether a state of
charge $(n_e,I)=(2,1)$ exists (see figure \ref{E1}b), and if there is one we
would like to determine its spin. Since the BPS algebra is real such a state
would be a marginally bound state of the W and the instanton.

\begin{figure}
\centerline{\epsfxsize=100mm\epsfbox{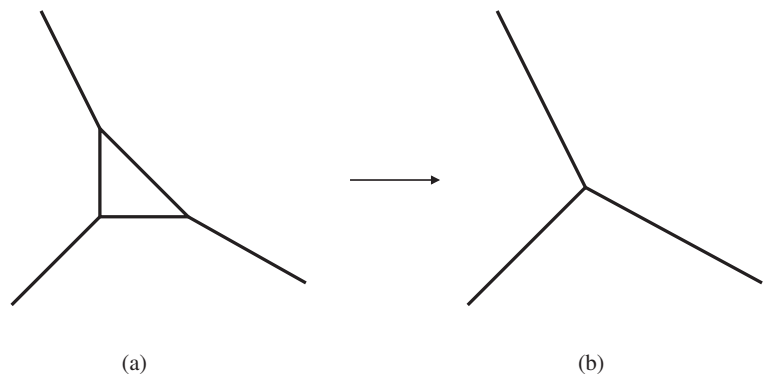}}
\medskip
\caption{A conformal theory singularity.}
\label{confsing}
\end{figure}

In addition to describing the prepotential and the BPS spectrum, the web
provides a picture of the different singularities in moduli space. A flop
transition is shown in figure (\ref{flop}). The mass of the hyper is
proportional to the finite segment. The transition point (figure \ref{flop}b)
has two intersecting lines creating a 4 junction. Note that this transition
between the two Kahler cones is natural in the web. Figure (\ref{Wsing}) shows a
singularity where a vector becomes massless enhancing the local gauge symmetry
as two parallel 5-branes coincide. Such a singularity is located on the boundary
of the extended Kahler cone. Figure (\ref{confsing}) shows a face shrinking to
zero size. Here a whole tower of BPS states becomes massless and we obtain a
conformal theory. This singularity is again located at the boundary of the
extended Kahler cone. These phase transitions and their relation to black hole
physics were also studied in \cite{Chou}.

Webs supply the Seiberg-Witten curve for a compactified 5d theory
\cite{BK5vM,BISTY} through a lift to M theory, in the same geometrical way that
was found in 4d \cite{Witten4vM}. The curve can be read from the corresponding
grid diagram. Integrable systems give another approach to 5d field theories
\cite{Nekrasov,LawNek,GGM,MarMir}. There, one considers formally the ''area"
differential $dx \wedge dy=d\lambda_{SW}$ to be the fundamental two form of a
dynamical system $dp
\wedge dq$, except that the area differential is complex rather than real.
\newpage
\section{Webs and toric geometry}
\label{toric}
Let us describe the correspondence between webs and Calabi-Yau spaces with a
shrinking 4-cycle \cite{LV}.
\begin{tabbing}
\hspace{3in} \= \kill
\underline{Web} \> \underline{Calabi-Yau} \\
$\bullet$ Web \>  Moment map
\end{tabbing}
It was found that webs correspond to spaces, $X$ that have a representation as a
toric variety. A $d$ complex dimensional toric variety is acted upon by the
algebraic torus, $\IC^{*d},\IC^*=\IC\backslash \{ 0\}$. It is defined by a fan
$\Delta \subset N$ where $N$ is a $d$ (real) dimensional vector space, and
$\Delta$ is collection of cones that constitutes the required combinatorial data
\cite{Fulton}. $N$ is the space of the grid diagram, and it seems that any grid
diagram can be mapped to a fan. The dual space to $N$ is called $M$. The web is
the image of the ''moment map" $X \to M$.
\begin{tabbing}
\hspace{3in} \= \kill
$\bullet$ Compact faces \> The shrinking 4-cycle.
\end{tabbing}
The 4-cycle is built from the planar face by ''fibering" a torus, $T^2$ above
each point.
\begin{tabbing}
\hspace{3in} \= \kill
$\bullet$ A \pq edge \> Locus of degeneration of a \pq \\
                        \> cycle of the torus.\\
\end{tabbing}
This ''fibration" has singularities where a \pq cycle on the torus shrinks.
These are the \pq edges in the web.
\begin{tabbing}
\hspace{3in} \= \kill
$\bullet$ External legs \> The normal bundle to the 4-cycle.\\
\end{tabbing}
To give a local description of the CY near a 4-cycle we still need to know the
structure of its normal bundle. Since the CY has $d=3$ the compact faces that
have $T^2$ ''fibered" over them, are already surfaces of degeneration of a cycle
in $T^3$. So on each face there is a cycle of $T^3$ that degenerates, and on
each edge actually two cycles degenerate.
\begin{tabbing}
\hspace{3in} \= \kill
$\bullet$The vertex condition \> The CY condition $c_1=0$\\
 $\bullet$ Triple intersections  of \> The CY triple intersection form\\
           faces/edges \>                on 4-cycles/2-cycles.\\
\end{tabbing}
 In addition, the vertex condition, and the intersections of the web have CY
parallels. The Seiberg - Witten curve, mentioned at the end of the previous
paragraph, seems to be related to the CY through mirror symmetry: it was found
that the mirror to a CY with a shrinking 4-cycle includes a Riemann surface
which depends only on the local 4-cycle and not on the surrounding CY
\cite{KKV}.
\section{BPS spectrum and Curves}
\label{BPSspec}
Having the mapping from the web to the Calabi-Yau space, we can use the results
of \cite{Witten_phase} to determine the BPS spectrum.
\begin{tabbing}
\hspace{3in} \= \kill
\underline{Web} \> \underline{Calabi-Yau} \\
$\bullet$ String web \> Curve
\end{tabbing}
 In the CY picture a BPS state comes from a membrane wrapping a curve. The moment map will send this
curve to a string web (assuming the curve is toric as well). A marginally bound
state exists when the curve is irreducible \cite{Witten_phase}. This corresponds
to the string web being irreducible.
\begin{tabbing}
\hspace{3in} \= \kill
$\bullet$ Irreducible string web \> Irreducible curve
\end{tabbing}

For {\it zero genus curves} the highest spin, $j$, of the BPS state is
determined by the dimension of its moduli space ${\cal M}$ \cite{Witten_phase}
\be
j=dim_{\IC}{\cal M}/2
\label{j}
\ee
Let us recall the derivation of this result. A massive BPS state in 5d
transforms under $SO(4)=SU(2)_p \times SU(2)_b$. Under this group the 8
supercharges transform as $2({\bf \half})_p \oplus 2({\bf \half})_b$. The BPS
state preserves 4 of the 8 supercharges. Without loss of generality we can take
them to be $2({\bf \half})_p$, while the broken ones are $2({\bf
\half})_b$. The 4 broken supercharges act on the state to produce half a
hypermultiplet $H_0=2({\bf 0})_b \oplus ({\bf 1/2})_b$. To get the total
representation we have to quantize the quantum mechanics on the moduli space of
the curve and tensor it with $H_0$.

The supersymmetric quantum mechanics takes place on the moduli space $\cal M$
which is a Kahler manifold. The Hilbert space is the collection of
\pq differential forms on $\cal M$, and the 4 supercharges are represented by
$\partial,\bar{\partial},\partial ^*,\bar{\partial}^*$.
For zero genus curves the fermions are simply in the tangent bundle of the
moduli space (as can be seen from the zero mode counting).
Supersymmetric ground states exist - they are the harmonic forms. These states
constitute a representation of $SU(2)_p$ in the following way: denote by
$J_3,J_+,J_-$ the standard $SU(2)$ generators. $J_3$ acts by multiplying a \pq
harmonic form by $(p+q-dim_{\IC}{\cal M})/2$, while $J_+,J_-$ act by a wedge
product or contraction with the Kahler form. The harmonic forms split into a
number of $SU(2)$ representations, the highest of which is (\ref{j}).

To map this discussion to the webs we recall that the image of a curve is a
string web. Every complex modulus of the curve is a real modulus, or a
deformation mode, since all phases are lost in the transition.
\begin{tabbing}
\hspace{3in} \= \kill
$\bullet$ Deformation mode \> Complex modulus
\end{tabbing}
The number of deformations, or bosonic zero modes, of a finite string web,
$n_{BZM}$ is given by $n_{BZM}=\#$(grid-points)$-1$ (ignoring the 4
translations), where we use the number of points in the corresponding grid
diagram, including boundary points. In terms of the web $n_{BZM}=F_{int}+n_X-1$
where $F_{int}$ is the number of (possible) internal faces, and $n_X$ is the
number of external legs \cite{AHK}. To see that imagine that the external legs
were infinite and add up the local modes $n_L=F_{int}$, the global modes
$n_G=n_X-3$ (which are also local for finite external legs) and 2 translation
modes.

When we consider non-zero genus curves, there are more fermions, and the bundle
of fermions is larger than the tangent bundle. To find the fermionic zero modes
of webs one should solve the equations given in \cite{BergmanKol}.
Alternatively, a short cut can be used. There are $n_X+F_{int}-1$ bosonic zero
modes, plus 4 space time translations. Whereas the ``supports" of $n_X-1$ modes
are preserved only by the 4 globally preserved supersymmetries, the  ``supports"
of the $F_{int}$ modes are preserved by 8 supersymmetries. The reason is that
these modes are supported on internal faces, and do not include the 5-brane
boundary. Finally we should add the 4 zero modes which are the partners of space
time translations. Summing all the contributions we find the number of fermionic
zero modes to be
\be
n_{FZM}=8F_{int}+4n_X.
\ee

Lifting to M theory, the string web turns into an open membrane ending on a
5-brane, and the 5d theory acquires a compact radius. Since the genus of the
smooth membrane is $g=F_{int}$ and the number of holes is $b=n_X$, we get
\bea
n_{BZM} = 2(g+b-1) \nonumber \\ n_{FZM}=8g+4b
\eea
This generalizes the results of \cite{HennYi,Mikhailov} that found that a disk
shaped membrane results in a hyper and a cylinder results in a vector.
\section{Examples}
\label{examples}
\subsection{$E_0$ or ${\bf P}^2$}
\begin{figure}
\centerline{\epsfxsize=100mm\epsfbox{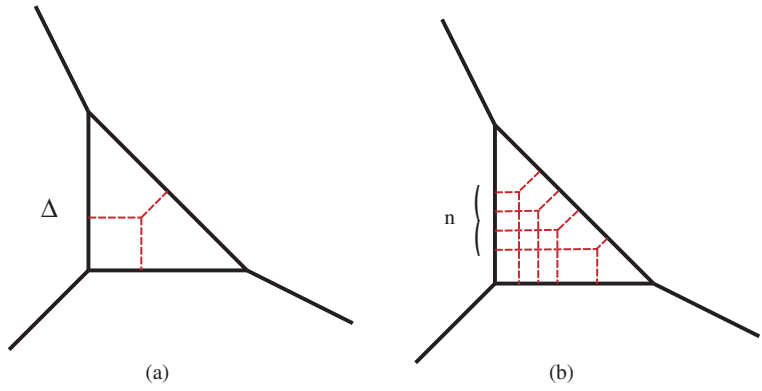}}
\medskip
\caption{The $E_0$ theory. (a) The BPS state $\Delta$. (b) A bound state of $n$ $\Delta$'s.}
\label{E0}
\end{figure}

\begin{figure}
\centerline{\epsfxsize=100mm\epsfbox{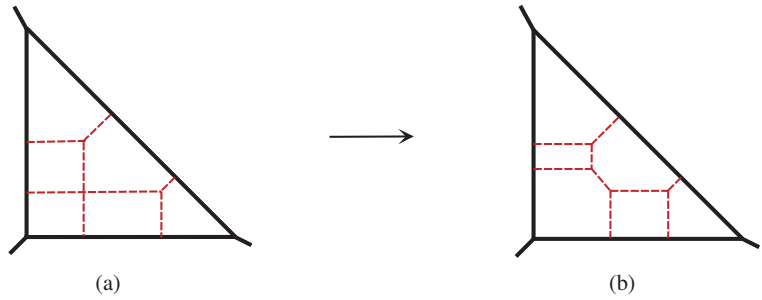}}
\medskip
\caption{ The BPS state made out of 2 $\Delta$'s. (a) A reducible configuration.
 (b) A general irreducible configuration.}
\label{E0-2}
\end{figure}

Consider the $E_0$ theory. The relevant web is shown in figure (\ref{E0}a). The
theory has a single simple BPS state, which we denote by $\Delta$, shown as the
dashed line. Consider a state made out of two $\Delta$'s (figure (\ref{E0-2}a).
Although this string web is reducible, a general deformation of it is
irreducible (figure (\ref{E0-2}b), so there is a marginally bound state.
Generalizing to $n$ $\Delta$ particles (\ref{E0}b) we see that a bound state
exists, and counting deformations
\be
n_{BZM}+1 = F = (n+2)(n+1)/2.
\label{defE0}
\ee
In particular, since the $\Delta$ does not have internal faces, we find that
\underline{$\Delta$ has $j=1$}. Note that in general the number of fermionic
zero modes of a bound state is higher than the sum of contributions from the
individual constituents.

The corresponding toric geometry is a shrinking ${\bf P}^2$. The same
computation can be repeated in elementary geometrical methods. $\Delta$
correspond to a line, ${\bf P}^1$. A bound state of $n$ $\Delta$'s is described
by a homogeneous polynomial of degree $n$ in the homogeneous coordinates of
${\bf P}^2$. Counting the number of coefficients for such a polynomial we find
that
\be
{\rm dim}_{\IC}{\cal M}+1=(n+2)(n+1)/2.
\ee
Comparing it with the previous result (\ref{defE0}) we verify that $n_{BZM}=
{\rm dim}_{\IC}{\cal M}$.

\subsection{$E_1$ or ${\bf P}^1 \times {\bf P}^1$}
\begin{figure}
\centerline{\epsfxsize=100mm\epsfbox{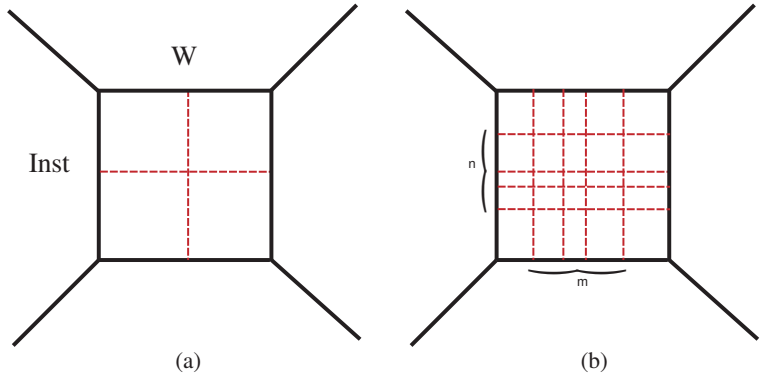}}
\medskip
\caption{The $E_1$ theory. (a) The W and the instanton. (b) A bound state of
 $m$ W's and $n$ instantons.}
\label{E1}
\end{figure}

\begin{figure}
\centerline{\epsfxsize=50mm\epsfbox{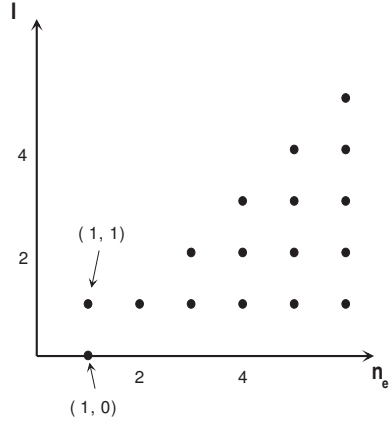}}
\medskip
\caption{The sites on the $E_1$ charge lattice which are occupied by BPS states.
 $n_e$ is the electric charge, $I$ is the instanton number.}
\label{E1charges}
\end{figure}

Consider the theory of pure $SU(2)$ gauge symmetry, a deformation of the $E_1$
fixed point (figure \ref{E1}). This theory was already mentioned at the end of
section \ref{WebRev}. Consider a state made out of $m$ W's and $n$ instantons
(figure \ref{E1}b). A bound state exists whenever $m,n>0$. The sites which are
certainly occupied on the $(n_e,I)$ charge lattice are shown in figure
(\ref{E1charges}). The number of bosonic zero modes is
\be
n_{BZM}+1=F=(m+1)(n+1)
\ee
In particular the bound state of the W and the instanton (figure \ref{E1}a) has
$j=3/2$.

The corresponding toric 4 cycle is ${\bf P}^1 \times {\bf P}^1$. The curve
spectrum on it consists of bihomogeneous curves of bidegree $(m,n)$. They are
indeed irreducible when $m,n>0$, and the dimension of the moduli space is ${\rm
dim}_{\IC}{\cal M}+1=(m+1)(n+1)$.

\subsection{$\tilde{E}_1$ or ${\bf F}_1$}
\begin{figure}
\centerline{\epsfxsize=100mm\epsfbox{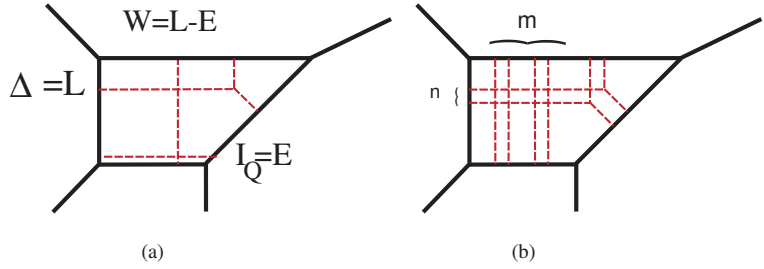}}
\medskip
\caption{The $\tilde{E}_1$ theory. (a) The W, the instantonic quark $I_Q$ and $\Delta$. (b) A bound state of
 $m$ W's and $n$ $\Delta$'s.}
\label{E1tilde}
\end{figure}

\begin{figure}
\centerline{\epsfxsize=50mm\epsfbox{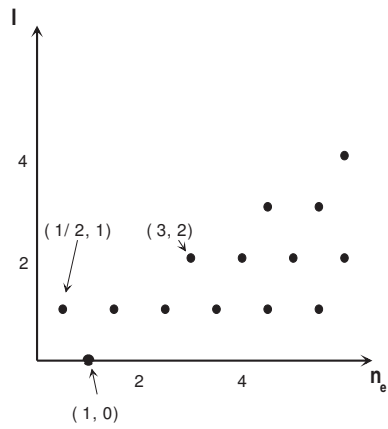}}
\medskip
\caption{The sites on the $\tilde{E}_1$ charge lattice which are occupied by BPS states.}
\label{E1tildecharges}
\end{figure}

Next we consider the theory of pure $SU(2)$ with a theta angle $\theta /\pi=1$
(mod 2) (figure \ref{E1tilde}). It is a deformation of the $\tilde{E}_1$ fixed
point. It has two simple BPS states, the W and the instantonic quark, $I_Q$,
which is a strip (we consider now only positive $m_0=1/g_0^2$). The $(n_e,I)$
charges are $(1,0)$ for the W and $(1/2,1)$ for $I_Q$. These two have a bound
state $\Delta$ with charge $(3/2,1)$. Consider a state made out of $m$ W's and
$n$ $\Delta$'s. We see that a bound state exists when $n>0$. The spectrum in
charge space is shown in figure (\ref{E1tildecharges}). The number of bosonic
zero modes is
\be
n_{BZM} +1={(m+n+2)(m+n+1)-m(m+1) \over 2}
\ee

The corresponding geometry is ${\bf F}_1$, the blow-up of ${\bf P}^2$ at one
point, $p$. $I_Q$ is the exceptional divisor, $E$. $\Delta$ is the line, $L$. W
corresponds to a line that passes through $p$, and belongs to the homology class
$L-E$. To see that it is the correct homology class we have to know the
intersection algebra on ${\bf F}_1$: $L \cdot L=1, L\cdot E=0, E \cdot E=-1$, so
indeed $(L-E)\cdot E=1$. To identify the curve spectrum we will use the fact
that the intersection number of curves (belonging to different classes) is
non-negative. Consider the class $m W+n \Delta=(m+n)L-m E$. From intersecting it
with $E$ we get $m \ge 0$, while from intersecting it with $L-E$ we get $n \ge
0$. The case $n=0$ is a curve of degree $m$ passing $m$ times through $p$ and is
clearly reducible. So the curve spectrum agrees. To find the dimension of moduli
space we have to count curves of degree $m+n$ which pass through a specific
point $m$ times. Imposing that the first $m$ derivatives vanish at $p$ we indeed
get ${\rm dim}_{\IC}{\cal M}+1=[(m+n+2)(m+n+1)-m(m+1)]/2$.

\newpage
\begin{center}
\large{ACKNOWLEDGEMENTS}
\end{center}
We thank O. Aharony, P. Biran, A. Hanany, A. Lawrence, A. Rajaraman, L. Susskind
and  C. Vafa. BK and JR are supported by NSF grant PHY-9219345.

\bibliography{5dspec}
\bibliographystyle{utphys}

\end{document}